# SQUID-based instrumentation for ultra-low-field MRI

Vadim S Zotev, Andrei N Matlashov, Petr L Volegov, Algis V Urbaitis, Michelle A Espy and Robert H Kraus, Jr[1]

Los Alamos National Laboratory, Group of Applied Modern Physics,
MS D454, Los Alamos, NM 87545, USA

E-mail: vzotev@lanl.gov (V S Zotev)

**Abstract**
Magnetic resonance imaging at ultra-low fields (ULF MRI) is a promising new imaging method that uses SQUID sensors to measure the spatially encoded precession of pre-polarized nuclear spin populations at a microtesla-range measurement field. In this work, a seven-channel SQUID system designed for simultaneous 3D ULF MRI and magnetoencephalography (MEG) is described. The system includes seven second-order SQUID gradiometers characterized by magnetic field resolutions of 1.2 – 2.8 fT/√Hz. It is also equipped with five sets of coils for 3D Fourier imaging with pre-polarization. Essential technical details of the design are discussed. The system's ULF MRI performance is demonstrated by multi-channel 3D images of a preserved sheep brain acquired at 46 microtesla measurement field with pre-polarization at 40 mT. The imaging resolution is 2.5 mm × 2.5 mm × 5 mm. The ULF MRI images are compared to images of the same brain acquired using conventional high-field MRI. Different ways to improve imaging SNR are discussed.

## 1. Introduction

Magnetic resonance imaging at ultra-low fields (ULF MRI) attracted considerable attention in recent years, both among superconductivity researchers and within MRI community. ULF MRI is an interesting new application of SQUID sensors and possible alternative to conventional high-field MRI. In this imaging method [1-10], nuclear spin population in a sample is pre-polarized [11] by a relatively strong (up to 0.1 T) magnetic field, and spin precession is encoded and detected at an ultra-low (<150 µT) measurement field after the pre-polarizing field is removed [1-10]. The ULF MRI signals are measured by SQUID sensors [12-16], commonly with untuned input circuit, that act as frequency-independent flux-to-voltage converters. High sensitivity of SQUIDs partially compensates for MRI signal reduction due to relatively low sample pre-polarization.

Magnetic resonance imaging at ULF has several advantages over conventional high-field MRI. Spatial resolution of MRI is ultimately determined by the NMR linewidth that depends on absolute field inhomogeneity. Because microtesla-range fields of modest relative homogeneity are highly homogeneous on the absolute scale, very narrow NMR lines with high SNR are achieved [1,5,9]. Thus, measurement field and encoding gradients for ULF MRI can be generated by simple and inexpensive coil systems [2,8]. Advantages of ULF MRI also include minimized susceptibility artifacts [1], enhanced $T_1$ contrast [17], and possibility of imaging in the presence of metal [5,18]. The main limitation of present-day ULF MRI in comparison to high-field MRI is its lower signal-to-noise ratio. The SNR can be improved through the use of stronger pre-polarizing fields, minimization of system noise, and optimization of pick-up coil geometry. Image distortions due to concomitant gradients are more pronounced at ULF and need to be corrected in practice [19,20].

Unlike conventional high-field imaging, ULF MRI is compatible with SQUID-based techniques for biomagnetic measurements [21] such as magnetoencephalography (MEG) [22] and magnetocardiography (MCG) [23]. It has been shown that ULF NMR signals from a human body can be measured simultaneously with MEG [6] or MCG [7] signals using the same SQUID sensor. Combining MEG and ULF MRI capabilities in a single instrument is particularly promising, because it will eliminate the need for MEG/MRI co-registraion and allow simultaneous functional (MEG) and anatomical (ULF MRI) imaging of the human brain. Because MEG systems typically include many channels, parallel imaging techniques developed in high-field MRI, such as SENSE [24] and PILS [25], can be readily applied at ULF to improve image quality or increase imaging speed.

Recently, we developed a seven-channel SQUID system specially designed for both ULF MRI and MEG [8]. The system was initially used to perform auditory MEG measurements and acquire a multi-channel 2D image of a human hand [8]. In the present work, we describe modification of this system for 3D ULF MRI. We discuss important technical details of the design and demonstrate performance of our system by acquiring multi-channel 3D images of a preserved sheep brain.

## 2. Instrumentation

### 2.1 Measurement system

The system we have developed for ULF MRI and MEG includes seven measurement channels (figure 1a). Each channel consists of a second-order axial gradiometer and a SQUID assembly (figure 1b). The gradiometers have 37 mm diameter and 60 mm baseline. We have selected this pick-up coil diameter (which is greater than in typical MEG instruments) to improve magnetic field resolution and imaging depth essential for ULF MRI while largely preserving the coils' ability to localize MEG sources. The gradiometers have been hand wound on G-10 fiberglass formers using 0.127 mm thick niobium wire. They are placed parallel to one another with one gradiometer in the middle and six others surrounding it in a hexagonal pattern shown in figure 1a. The center-to-center spacing of the neighboring pick-up coils is 45 mm. The seven channels are installed inside a flat-bottom fiberglass liquid helium cryostat (available as Model LH-14.5-NTE from Cryoton Co. Ltd., www.cryoton.webzone.ru).

Each SQUID assembly consists of a SQUID sensor and a cryoswitch (available as Model CE2 blue and Model SW1, respectively, from Supracon AG, www.supracon.com). The sensor's input coil inductance $L_i$ is 420 nH, and the mutual inductance $M_i$ between the input coil and the SQUID is 8.1 nH. The inductance $L_p$ of our second-order gradiometers was estimated to be ≈500 nH. The cryoswitch, included in the input circuit between the gradiometer and the SQUID as illustrated in figure 1b, is superconductive at 4 K, but becomes resistive when its heater is activated. The resistance is 50 ohm, and the switching time is about 5 microseconds. The cryoswitch is used to protect the SQUID from transients caused by rapid switching of the pre-polarizing field in ULF MRI experiments. The SQUID assembly is enclosed in a lead shield and installed 12 cm above the gradiometer.

Figure 2 exhibits noise spectra of the system operated inside a two-layer magnetically shielded room. Magnetic flux noise spectral density was measured to be about 5 $\mu\Phi_0/\sqrt{Hz}$ at 1 kHz for channel 1. Because magnetic flux in the SQUID is equal to magnetic flux in the pick-up coil times $M_i/(L_p+L_i)$, magnetic field noise spectral density, referred to the pick-up coil, is 1.2 fT/√Hz for channel 1. For the surrounding channels, the noise spectral densities are 2.5-2.8 fT/√Hz at 1 kHz. This increase in the noise level for the outside channels is mainly due to Johnson noise originating from the thermal shield (consisting of 1.6 mm-diameter aluminum rods) between the vertical walls of the cryostat.

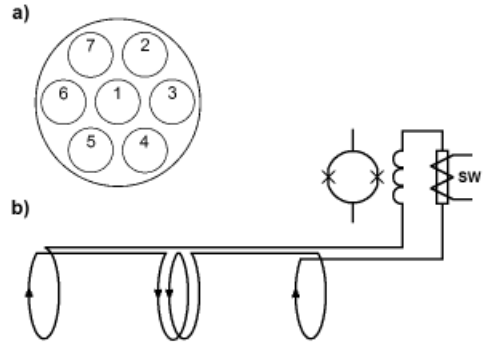

**Figure 1**. a) Positions of the seven channels inside the cryostat; b) schematic of one channel.

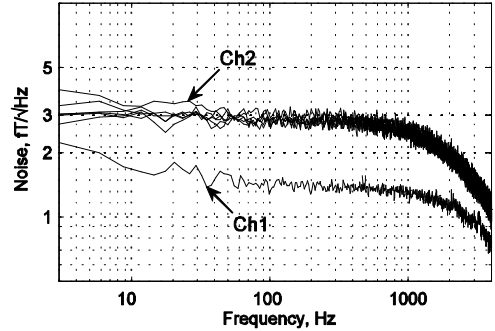

**Figure 2**. Noise spectra of the seven channels.

The noise spectra remain essentially flat down to frequencies of a few hertz. The interference-free operation is ensured by careful organization of electrical connections: all equipment inside the shielded room is powered from batteries and connected to outside electronics by fiber optic cables. The signal roll-off for frequencies in the kilohertz range (figure 2) is a characteristic of the data acquisition board used (available as Model ADCXF-4412 from Research Electronics Development, Inc., www.redhitech.com). We have selected this 24 bit ADC with 16 kHz sampling frequency, because it is equipped with a fiber optic interface. This allows the data acquisition board, placed inside the shielded room, to be electrically isolated from the computer outside. The bandwidth of the SQUID electronics is 50 kHz.

### 2.2 Coil system

Schematic layout of the coil system for 3D ULF MRI is exhibited in figure 3. The system includes five sets of coils. A pair of round Helmholtz coils, 120 cm in diameter, provides microtesla-range measurement field $B_m$ along the Z axis. The measurement field strength is 46 µT at 750 mA. Three sets of coils generate three gradients for 3D Fourier imaging. The longitudinal gradient $G_z=dB_z/dz$ is produced by two 80 cm square Maxwell coils. The magnitude of this gradient at 1 A current is 120 µT/m. A set of eight rectangular coils on two 48 cm × 96 cm frames orthogonal to the X axis creates the transverse gradient $G_x=dB_z/dx$. The $G_x$ strength is 80 µT/m at 1 A. The second transverse gradient, $G_y=dB_z/dy$, is generated by a set of four rectangular coils on two 62 cm ×

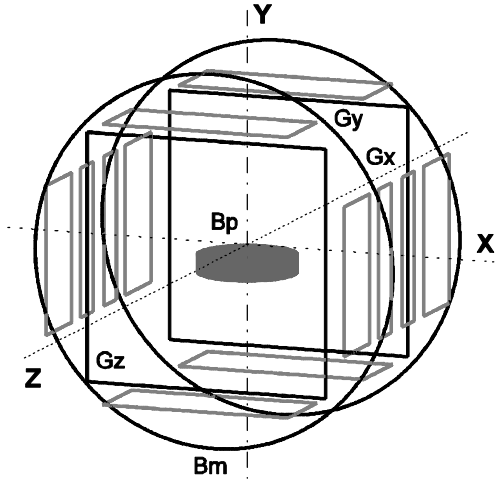

**Figure 3**. Schematic layout of the coil system for 3D ULF MRI.

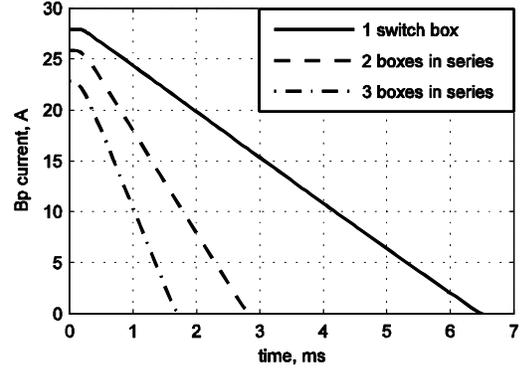

**Figure 4**. Switch-off profile of the pre-polarizing current.

96 cm frames orthogonal to the Y axis. The magnitude of $G_y$ is 140 µT/m at 1 A. Each of these coil sets is symmetric with respect to the center of the system. The system as a whole should be carefully centered inside the shielded room.

The pre-polarizing field $B_p$ in our system is three orders of magnitude stronger than the measurement field $B_m$. It is produced by a cylindrical coil with 32 cm outside diameter, 18 cm inside diameter, and 7 cm thickness. The number of turns is 800, and the coil inductance is 100 mH. To reduce eddy currents induced inside the coil by rapid switching of the pre-polarizing field, a special wire was used (available as Twistite brand from MWS Wire Industries, www.mwswire.com). It was manufactured by twisting together 24 strands of copper wire (AWG 28) with heat-resistant polyimide insulation. We selected this type of wire, because, in our experience, the commonly available litz wire with thin polyurethane insulation could be easily damaged by accidental coil overheating. The $B_p$ coil is immersed in liquid nitrogen bath. When it is cooled down from room temperature to 77 K, its resistance decreases 6 times, and heat dissipation drops accordingly. The sample is placed at the center of the imaging coil system, and the $B_p$ coil is positioned underneath the sample. The y-component of the $B_p$ field varies between 40 mT and 50 mT at 28 A along the vertical dimension of the sample space.

For samples with relatively short longitudinal relaxation time $T_1$, it is important to turn off the pre-polarizing field quickly to prevent loss in polarization. In our system, the $B_p$ field can be cut to zero in a few milliseconds as shown in figure 4. This is accomplished with switch boxes consisting of optoMOS relays (available as Model CPC1918J from CLARE, www.clare.com). A switch box is connected between the $B_p$ coil and a voltage supply that includes several batteries. When a control signal is turned off, a MOSFET inside a relay begins to close, and a high voltage spike is generated across it by the coil inductance. Because this high voltage (typically in the kilovolt range) exceeds the MOSFET's breakdown voltage $V_{BR}$, an avalanche breakdown occurs, during which voltage across the MOSFET remains constant and equal to $V_{BR}$. The current therefore drops at a constant rate, $dI/dt = -(V_{BR} - V_S)/L$, where $V_S$ is the supply voltage and $L$ is the coil inductance. The inductive energy is dissipated by the MOSFET during this process. If several optoMOS relays are connected in series, their breakdown voltages add up, and the switching rate is roughly proportional to the number of relays. In contrast, parallel connection of the relays does not change the switch-off rate visibly. Our typical switch box includes four series-connected sections each consisting of four parallel relays (to increase current limit), mounted on a heat sink with either air or water cooling. With $V_{BR} \sim 120$ V for the type of relay used, $V_S = 24$ V, and $L = 100$ mH, the expected switching rate is 4.6 A/ms. This agrees well with the experimental data in figure 4, where 28 A current is turned off linearly in about 6 ms using one switch box. This time can be reduced further if switch boxes are connected in series as demonstrated in figure 4.

Each coil in the system is characterized by some distributed capacitance. Because of this, switching fields and gradients can induce electromagnetic oscillations within a coil even if the coil itself is disconnected from any external circuit. For example, switching gradients have been observed to induce parasitic currents in the $B_p$ coil, which were picked up by the SQUIDs and appeared as bright spots scattered across ULF MRI images. To ensure quick dissipation of such currents, each coil in our system is shunted with a large resistor, located as close to the coil as possible. The optimum value of this resistor has been empirically determined to be approximately 1000 times higher than the coil's own resistance.

### 2.3 Imaging procedure

The ULF MRI images reported in the next section were acquired according to the imaging protocol shown in figure 5. Each imaging step begins with pre-polarization of the sample by the field $B_p$ during time $t_p$ that should be at least as long as the relaxation time $T_1$. The pre-polarizing field is then turned off rapidly, and the measurement field $B_m$ is applied perpendicular to the direction of $B_p$. The application of $B_m$ induces spin precession. In our previous work [8], the measurement field was kept constant throughout the experiment, and spin precession was induced by the removal of $B_p$. The protocol in figure 5 has two advantages. First, there is no need to remove the $B_p$ field non-adiabatically (i.e. fast compared to the period of Larmor precession in the field $B_m$). Application of the $B_m$ field does not need to be non-adiabatic either, because spin precession starts as soon as a tiny fraction of $B_m$ appears perpendicular to the magnetization vector. Second, variations in the switching time, caused, for exam-ple, by drifting $B_p$ current, do not introduce spurious phase shifts between different measurements. The gradient coils, as well as the $B_m$ coils, are disconnected during the pre-polarization to minimize transients. After the measurement field is applied, imaging is performed according to the standard 3D Fourier imaging protocol with gradient echo. Phase encoding is carried out during time $t_g$ with two gradients, $G_z$ and $G_y$, and gradient echo is created by reversal of the frequency encoding gradient $G_x$ (figure 5). We have found that reversal of the measurement field $B_m$ simultaneously with the gradient $G_x$ further improves imaging SNR, but this technique was not used in the described experiment.

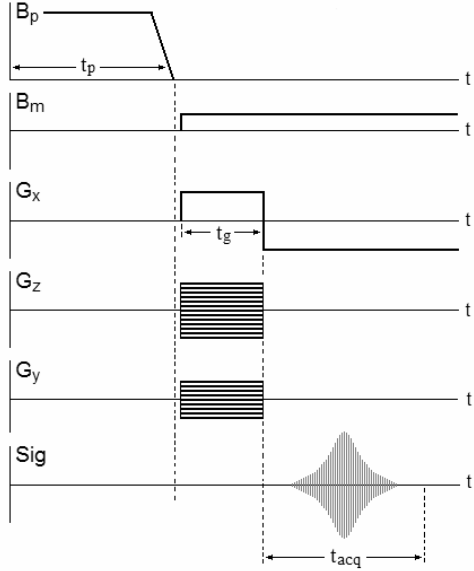

**Figure 5**. 3D Fourier imaging protocol with gradient echo in ULF MRI.

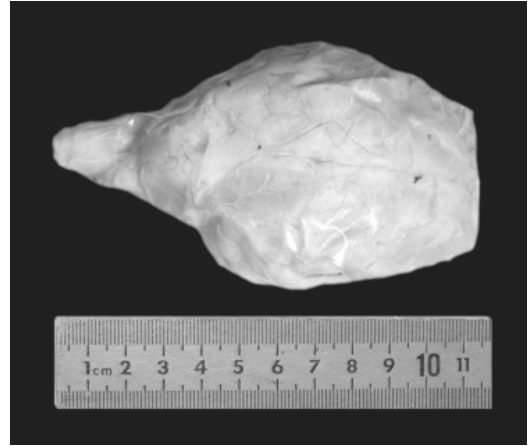

**Figure 6**. Photograph of the preserved sheep brain used for imaging.

In the sheep brain imaging experiment, the pre-polarization time was $t_p$=0.5 s, and the pre-polarizing field $B_p$ varied between 40 and 50 mT along the vertical dimension of the brain. The measurement field $B_m$≈46 µT corresponded to Larmor frequency of about 1940 Hz. We chose this frequency as the highest practical frequency within the range of our data acquisition board (figure 2). This allowed us to reduce the effect of concomitant gradients and to avoid magnetic noise induced by mechanical vibrations (resulting from $B_p$ pulses) below 1 kHz. The encoding and acquisition times were $t_g$=33 ms and $t_{acq}$=66 ms, respectively. The frequency encoding gradient $G_x$ had ±140 µT/m (±60 Hz/cm) values. The phase encoding gradient $G_z$ had limiting values ±140 µT/m (±60 Hz/cm) with 33 phase encoding steps. A total of 11 phase encoding steps were taken for the y-direction, with the limiting gradient values $G_y$= ±70 µT/m (±30 Hz/cm). Thus, each k-space scan included 363 measurements and required 4 minutes. It should be noted, however, that 80% of this time was used for pre-polarization. The described sequence provided 2.5 mm × 2.5 mm × 5 mm imaging resolution.

## 3. Results

To study the system's 3D ULF MRI performance in preparation for human brain imaging, we acquired 3D images of a preserved sheep brain (available from Fisher Science Education, www.fishersci.com). The brain is shown in figure 6. It is preserved in formaldehyde. The brain's longitudinal relaxation time $T_1$ was estimated to be about 300 ms. The mean relaxation time $T_2$ was measured to be approximately 40 ms at 46 µT field. High-field MRI measurements at 2 tesla field gave similar results. For comparison, the $T_2$ time of a living human brain varies between 80 ms and 100 ms (at high fields). This means that imaging the preserved brain is a more difficult task, because MRI signal drops faster with time.

Results of the brain imaging experiment at 46 microtesla measurement field are presented in figure 7. The imaging was performed according to the procedure described in the previous section. The sheep brain was placed horizontally under the bottom of the cryostat with its longest dimension along the line of channels 1, 3, and 6. Each image in figure 7 represents a 5 mm thick horizontal layer of the brain with the vertical position of the central plane specified by coordinate Y. The value Y=0 corresponds to the center of the gradient coil system. The bottom of the cryostat is located at Y≈25 mm. The in-plane resolution is 2.5 mm by 2.5 mm. The images were acquired simultaneously as parts of one 3D image, reconstructed from measured gradient echo signals by 3D Fourier transform. To improve signal-to-noise ratio for deep-lying layers of the brain, 42 scans of k-space were performed consecutively, and the images were averaged. The total imaging time was about 3 hours. The effect of concomitant gradients was corrected for all the ULF images in this paper according to the method of [20]. Each image in figure 7 is a composite image with sensitivity correction, computed from the seven individual-channel images using PILS method [25] with experimentally determined sensitivity maps of the channels. Sensitivity properties of the seven channels are discussed in more detail below. All the images in figure 7 were additionally interpolated.

Figure 8 exhibits images of the same sheep brain, acquired by conventional high-field MRI. The imaging was performed in a whole-body MRI scanner with 2 tesla superconducting magnet. The slice selection technique was employed. The horizontal slices were 2 mm thick with 5 mm distance between their central planes. The in-plane resolution was 1 mm by 0.5 mm. The spin echo time was $TE$=26 ms, with the repetition time of $TR$=1500 ms. Each image in figure 8 is an average of four measured images.

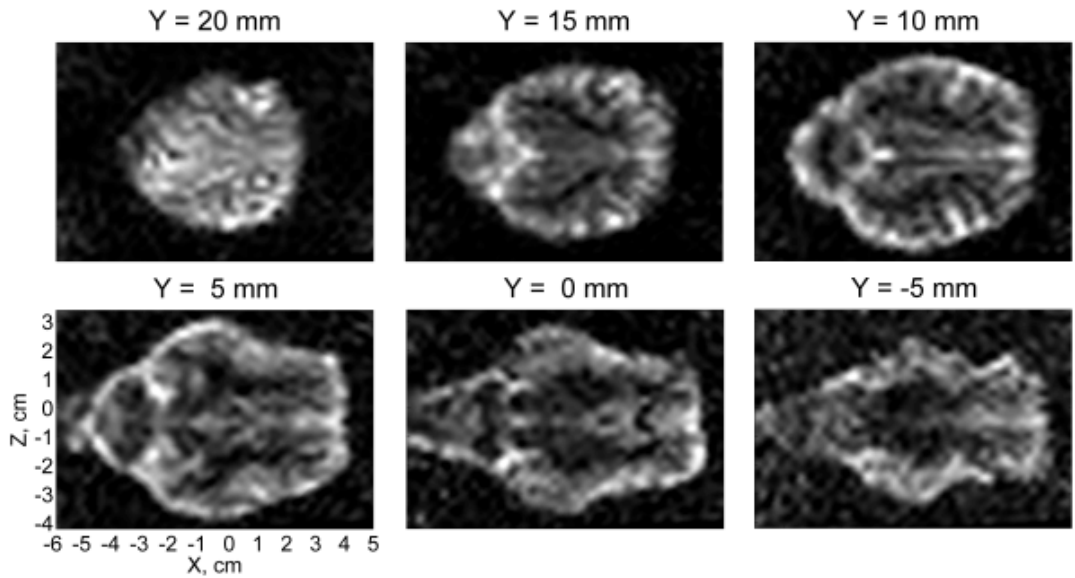

**Figure 7**. Images of the preserved sheep brain acquired by 3D Fourier ULF MRI at 46 µT field.

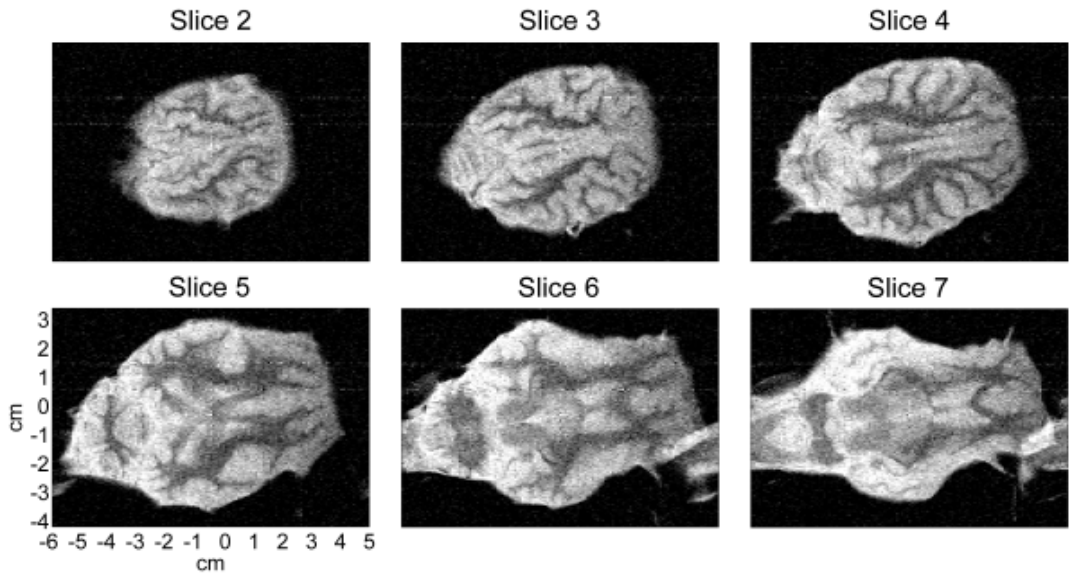

**Figure 8**. Images of the same brain obtained by high-field (2 tesla) MRI with slice selection.

Comparing results in figure 7 and figure 8, we conclude that the ULF images provide essentially the same anatomical information as the corresponding high-field images. Precise comparison, however, is difficult because of the differences in brain position, slice thickness, and in-plane resolution. Both sets of images are $T_2$-weighted images, with darker regions characterized by shorter $T_2$ times. The ULF images exhibit greater contrast, but this can be attributed to the fact that the gradient echo time in the ULF MRI experiment (counted from the moment $B_m$ field is applied) is about 70 ms as opposed to $TE$=26 ms in the high-field measurements.

To characterize the system sensitivity and perform sensitivity correction of images in figure 7, we acquired 3D sensitivity maps of the channels. The maps were obtained by imaging a thick, 23 cm in diameter, uniform water phantom with the same spatial resolution as that in the sheep brain experiment. A composite seven-channel sensitivity map (computed as a square root of the sum of squares of images from the seven individual channels) for a 5 mm thick layer with $Y$=20 mm is exhibited in figure 9. The map, originally distorted by concomitant gradient artifacts, is symmetric after this distortion has been corrected.

Figure 10 compares sensitivity maps for different image layers. It shows sensitivity profiles along the line of channels 1, 3, and 6 (which is close to X axis) for the six image layers with the same values of $Y$ as in figure 7. The single-peak curves ($S_1$) are sensitivities of channel 1, while the curves with three maxima ($S_c$) represent the composite seven-channel sensitivities. These curves, derived from the experimental sensitivity maps, were digitally smoothed to improve their appearance. According to figure 10, the peak sensitivity of channel 1 for the bottom image layer ($Y$= -5 mm) is 4 times lower than for the top layer ($Y$=20 mm). This decrease in sensitivity with the distance from the pick-up coils explains why extensive averaging is required to obtain good-quality images of the deeper layers of the brain. Figure 10 also shows how the use of multiple channels improves imaging field-of-view and signal-to-noise. While the sensitivity increase is rather modest at the center of the system (from 1% for the top to 18% for the bottom layer), the $S_c/S_1$ ratio grows considerably as the observation point moves from the center towards any of the six outside channels.

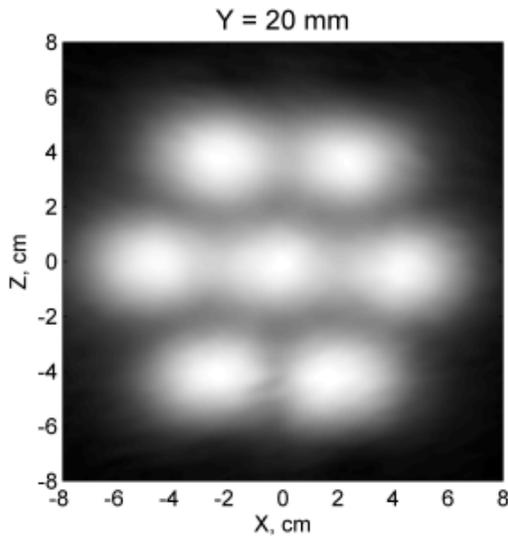

**Figure 9**. Composite seven-channel sensitivity map for one image layer.

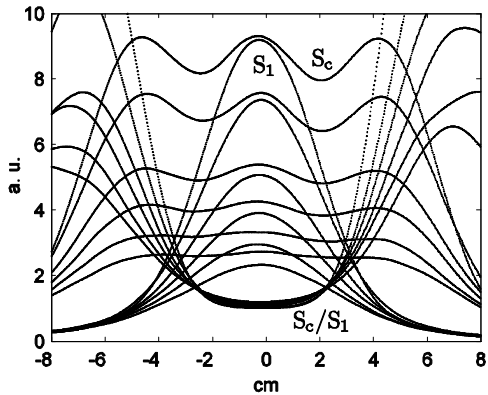

**Figure 10**. Sensitivity profiles ($S_1$ – channel 1, $S_c$ – composite) for six image layers.

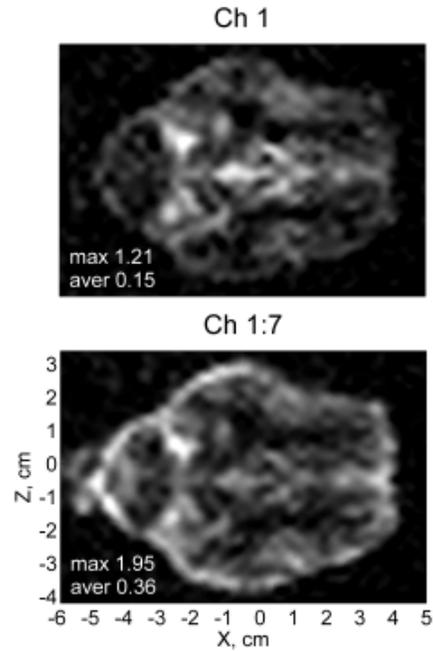

**Figure 11**. Comparison of sheep brain images (for $Y$=5 mm) acquired by one and seven channels.

Figure 11 demonstrates the image improvement through the use of multiple channels predicted in figure 10. The images of the same layer (with $Y$=5 mm) were calculated using data from one and seven channels, respectively. The seven-channel image was computed as a square root of the sum of squares without sensitivity correction. When compared to the single-channel image, it shows only minor (10%) improvement in the central brain region, but exhibits an average 3-fold increase in signal strength near the edges of the brain, where the sensitivity of channel 1 is low. This result, which is consistent with figure 10, demonstrates the benefit of multi-channel imaging even for those objects that are small enough and can be imaged by a single channel.

## 4. Conclusion

In this paper, we described a seven-channel SQUID system designed for both 3D ULF MRI and MEG. We also presented, for the first time, multi-channel 3D images of a preserved sheep brain acquired at an ultra-low field. The ULF MRI images reveal the same anatomical details as the brain images obtained by conventional high-field MRI. They suggest that our system, after proper modification, can be successfully used for human brain imaging. Our results also demonstrate that the presently available imaging SNR remains the most serious limitation in ULF MRI. Because of the insufficient SNR, extensive averaging is required to improve image quality for biological objects with relatively short $T_2$ times. The SNR of our system can be increased in at least three different ways.

First, one can reduce the cryostat noise by minimizing the Johnson noise contribution from the thermal shield. This can be achieved by replacing the thick aluminum rods with bundles of thin insulated wires. Such a modification would reduce noise levels of six channels surrounding channel 1 in our system by a factor of 2. The cryostat noise can be reduced further if a special low-noise super-insulation is used.

Second, one can increase the channels' sensitivity depth by using larger-diameter pick-up coils. For high-quality SQUIDs, magnetic flux noise is virtually independent of the pick-up coil inductance. This inductance is proportional to the coil radius, $L_p = cR$, and, for large $R$, exceeds the inductance $L_i$ of the SQUID input coil. The flux transfer coefficient becomes $M_i/(cR)$. The SNR, defined as a ratio of flux to (constant) flux noise, is thus proportional to $\Phi(R)/R$, where $\Phi(R)$ is magnetic flux penetrating the pick-up coil. For a uniform magnetic field, the flux is proportional to the coil area, so the SNR is a linear function of $R$. For a magnetic dipole at a distance $d$ below the pick-up coil, $\Phi(R)/R$ increases almost linearly with $R$ until it reaches maximum at $R_m = d/\sqrt{2}$, and then begins to decrease. Therefore, the maximum sensitivity depth is proportional to the pick-up coil radius $R$. This analysis shows that larger pick-up coils offer an advantage of higher imaging SNR. One can envision a combined MEG/ULF-MRI system with pick-up coils of two different diameters: the smaller pick-up coils could be used for MEG source localization, and the larger ones – for anatomical imaging. With our present cryostat, however, $R$ can only be increased 2-3 times at the expense of reduction in the number of imaging channels.

Third, one can increase the strength of the pre-polarizing field $B_p$. This approach is preferable, because it does not require rebuilding the cryostat or making new gradiometers. Also, there is no upper limit on SNR improvement that can be achieved. Strong $B_p$ fields, however, can only be generated if an efficient cooling system for the $B_p$ coils is developed. This task is further complicated in the case of human brain imaging by the need to safely accommodate a human head inside the system. Our new system of cooled $B_p$ coils for human brain imaging is currently being built. Because pre-polarization takes 80% of total imaging time in our experiments, improvements to the imaging protocol that make more efficient use of a sample's polarization would also be very beneficial.

Overall, our experimental results suggest that SQUID-based ULF MRI can potentially be as efficient as high-field MRI. Magnetoencephalography, when combined with ULF MRI, will become a more versatile, accurate, and appealing technique.

**Acknowledgments**

We gratefully acknowledge the support of the U.S. National Institutes of Health Grant R01-EB006456 and of the Los Alamos National Security, LLC, for the National Nuclear Security Administration of the U.S. Department of Energy Grant LDRD-20060312ER.

**References**

[1] McDermott R, Trabesinger A H, Mück M, Hahn E L, Pines A and Clarke J 2002 Liquid-state NMR and scalar couplings in microtesla magnetic fields *Science* **295** 2247-2249

[2] McDermott R, Lee S-K, ten Haken B, Trabesinger A H, Pines A and Clarke J 2004 Microtesla MRI with a superconducting quantum interference device *Proc. Nat. Acad. Sci.* **101** 7857-7861

[3] McDermott R, Kelso N, Lee S-K, Mößle M, Mück M, Myers W, ten Haken B, Seton H C, Trabesinger A H, Pines A and Clarke J 2004 SQUID-detected magnetic resonance imaging in microtesla magnetic fields *J. Low Temp. Phys.* **135** 793-821

[4] Mößle M, Myers W R, Lee S-K, Kelso N, Hatridge M, Pines A and Clarke J 2005 SQUID-detected in vivo MRI at microtesla magnetic fields *IEEE Trans. Appl. Supercond.* **15** 757-760

[5] Matlachov A N, Volegov P L, Espy M A, George J S and Kraus R H Jr 2004 SQUID detected NMR in microtesla magnetic fields *J. Magn. Reson.* **170** 1-7

[6] Volegov P, Matlachov A N, Espy M A, George J S and Kraus R H Jr 2004 Simultaneous magnetoencephalography and SQUID detected nuclear MR in microtesla magnetic fields *Magn. Reson. Med.* **52** 467-470

[7] Espy M A, Matlachov A N, Volegov P L, Mosher J C and Kraus R H Jr 2005 SQUID-based simultaneous detection of NMR and biomagnetic signals at ultra-low magnetic fields *IEEE Trans. Appl. Supercond.* **15** 635-639

[8] Zotev V S, Matlachov A N, Volegov P L, Sandin H J, Espy M A, Mosher J C, Urbaitis A V, Newman S G and Kraus R H Jr 2007 Multi-channel SQUID system for MEG and ultra-low-field MRI *IEEE Trans. Appl. Supercond.* **17** 839-842

[9] Burghoff M, Hartwig S and Trahms L 2005 Nuclear magnetic resonance in the nanotesla range *Appl. Phys. Lett.* **87** 054103

[10] Burghoff M, Hartwig S, Kilian W, Vorwerk A and Trahms L 2007 SQUID systems adapted to record nuclear magnetism in low magnetic fields *IEEE Trans. Appl. Supercond.* **17** 846-849

[11] Macovski A and Conolly S 1993 Novel approaches to low cost MRI *Magn. Reson. Med.* **30** 221-230

[12] Fan N Q, Heaney M B, Clarke J, Newitt D, Wald L L, Hahn E L, Bielecki A and Pines A 1989 Nuclear magnetic resonance with dc SQUID preamplifiers *IEEE Trans. Magn.* **25** 1193-1199

[13] Seton H C, Bussell D M, Hutchison J M S and Lurie D J 1995 Use of a DC SQUID receiver preamplifier in a low field MRI system *IEEE Trans. Appl. Supercond.* **5** 3218-3221

[14] Kumar S, Avrin W F and Whitecotton B R 1996 NMR of room temperature samples with a flux-locked dc SQUID *IEEE Trans. Magn.* **32** 5261-5264

[15] Schlenga K, McDermott R, Clarke J, de Souza R E, Wong-Foy A and Pines A 1999 Low-field magnetic resonance imaging with a high-$T_c$ dc superconducting quantum interference device *Appl. Phys. Lett.* **75** 3695-3697

[16] Greenberg Y S 1998 Application of superconducting quantum interference devices to nuclear magnetic resonance *Rev. Mod. Phys.* **70** 175-222

[17] Lee S-K, Mößle M, Myers W, Kelso N, Trabesinger A H, Pines A and Clarke J 2005 SQUID-detected MRI at 132 μT with T1-weighted contrast established at 10 μT-300mT *Magn. Reson. Med.* **53** 9-14

[18] Mößle M, Han S I, Myers W R, Lee S-K, Kelso N, Hatridge M, Pines A and Clarke J 2006 SQUID-detected microtesla MRI in the presence of metal *J. Magn. Reson.* **179** 146-151

[19] Volegov P L, Mosher J C, Espy M A and Kraus R H Jr 2005 On concomitant gadients in low-field MRI *J. Magn. Reson.* **175** 103-113

[20] Myers W R, Mößle M and Clarke J 2005 Correction of concomitant gradient artifacts in experimental microtesla MRI *J. Magn. Reson.* **177** 274-284

[21] Sternickel K and Braginski A I 2006 Biomagnetism using SQUIDs: status and perspectives *Supercond. Sci. Technol.* **19** S160-S171

[22] Hamalainen M, Hari R, Ilmoniemi R J, Knuutila J and Lounasmaa O V 1993 Magnetoencephalography – theory, instrumentation, and applications to noninvasive studies of the working human brain *Rev. Mod. Phys.* **65** 413-497

[23] Stroink G, Moshage W and Achenbach S 1998 Cardiomagnetism *Magnetism in Medicine*, Ed W Andra and H Nowak (Berlin:Wiley-VCH) pp 136-189

[24] Pruessmann K P, Weiger M, Scheidegger M B and Boesiger P 1999 SENSE: sensitivity encoding for fast MRI *Magn. Reson. Med.* **42** 952-962

[25] Griswold M A, Jacob P M, Nittka M, Goldfarb J W and Haase A 2000 Partially parallel imaging with localized sensitivities (PILS) *Magn. Reson. Med.* **44** 602-609